\documentclass{aims}

\def\typeofarticle{Research Article}
\def\currentvolume{3}
\def\currentissue{x}
\def\currentyear{2015}

\def\ppages{xxx--xxx}
\def\DOI{10.3934/ms.2015.x.xxx}
\def\Received{July 2015}
\def\Accepted{September 2015}
\def\Published{December 2015}

\usepackage[font=footnotesize]{subfig}
\usepackage{algpascal}
\usepackage[inline]{trackchanges}
\usepackage{cite}
\addeditor{JB}
\addeditor{AG}
\addeditor{DS}
\addeditor{CT}

\begin{document}

% \title{Comprehensive Simulation of Complex Random Memristive Networks for Real-Time Computing }
% \title{Modeling and Simulation of Complex Random Resistive Switch Networks}
\title{Computational Capacity and Energy Consumption of Complex Resistive Switch Networks}

% \title{The full title of your paper}

\author{%
  Jens B\"{u}rger \affil{1}\corrauth
  Alireza Goudarzi\affil{2}\corrauth
  Darko Stefanovic\affil{2}
  and
  Christof Teuscher\affil{1}
}

% \shortauthors is used in copyright information in the end of the paper
\shortauthors{author name}

\address{%
  \addr{\affilnum{1}}{Department of Electrical and Computer Engineering, Portland State University, Portland, OR, 97211, USA}
  \addr{\affilnum{2}}{Department of Computer Science, University of New Mexico, Albuquerque, NM, 87131, USA}}

% corresponding author
\corraddr{jb22@pdx.edu, alirezag@cs.unm.edu}

\begin{abstract}

Resistive switches are a class of emerging nanoelectronics devices that  exhibit a wide variety of switching characteristics  closely resembling behaviors of biological synapses. Assembled into random networks, such resistive switches produce emerging behaviors far more complex than that of individual devices. This was previously demonstrated in simulations that exploit information processing within these random networks to solve tasks that require nonlinear computation as well as memory. Physical assemblies of such networks manifest complex spatial structures and basic processing capabilities often related to biologically-inspired computing. We  model and simulate random resistive switch networks and analyze their computational capacities. We provide a detailed discussion of the relevant design parameters and establish the link to the physical assemblies by relating the modeling parameters to physical parameters. 
%We show how computational capacities grow with energy consumption and discuss a modular approach that can increase computational capacities using multiple independent networks with morphologies related to low-energy consumption. 
More globally connected networks and an increased network switching activity are means to increase the computational capacity linearly at the expense of exponentially growing energy consumption. We discuss a new modular approach that exhibits higher computational capacities and energy consumption growing linearly with the number of networks used.
The results show how to optimize the trade-off between computational capacity and energy efficiency and are relevant for the design and fabrication of novel computing architectures that harness random assemblies of emerging nanodevices. 
\vfill

\end{abstract}

\keywords{
\textbf{Memrisive Networks, Atomic Switch Networks, Network Model, Reservoir Computing}
}

\maketitle

% \section{Questions to be addressed...}

\section{Introduction}

% \note[CT]{feels a bit repetitive and somewhat disorganized} \newline
% \note[CT]{I think RC and memristors need to be introduced} \newline

% \note[JB]{I am thinking of the introduction covering 4 aspects as follows. The text might not fit perfectly anymore}\newline

% \note[JB]{1. Intro to the fabrication of ASN and why is it relevant to computation.} \newline

Emerging nanoscale resistive switches provide possible solutions for creating future computing architectures that are faster, less expensive and more energy-efficient by exploiting their intrinsic switching characteristics \cite{Crutchfield2010}. Recent progress in the fabrication of atomic switches \cite{Hasegawa2010}, as one class of emerging nanodevices, allows the random assembly of these devices into larger networks \cite{Avizienis2012, Sillin2013, Stieg2012}. In \cite{Demis2015} it was argued that the complexity of these networks resembles that of biological brains, in which the complex morphology and interactions between heterogeneous network elements are responsible for powerful and energy-efficient information processing \cite{Sporns2011}. Contrary to designed computation, where each device has a specified role, computation in random resistive switch networks does not rely on specific devices, but is encoded in the collective nonlinear dynamic switching behavior as a result of an applied input signal. 

% \note[JB]{2.Origin of the computation with respect to the device characteristics}\newline
Atomic switches, as well as memristive devices \cite{Chua1971, Strukov2008}, are history-dependent resistive switches. Application of a bias voltage can change the conductance of the device. 
%Based on the nonlinear relationship between an applied input bias and a resulting change in conductance, these devices offer the possibility to perform nonlinear transformations of the input signal. The resulting device dynamics are similar to that found in biological synapses \cite{Jo2010, Jo2011, Kim2015, Ohno2011, Sah2014}.
The nonlinear relationship between an applied input and a resulting conductance change in these devices perform nonlinear transformations of the input, similar to the dynamics found in biological synapses \cite{Jo2010, Kim2015, Ohno2011, Sah2014}.

% \note[JB]{3.Existing results}\newline
Harnessing the intrinsic nonlinear characteristics of these emerging nanodevices assembled in random structures has been shown for the memristor \cite{Burger2013, Burger2015, Kulkarni2012} as well as for the atomic switch \cite{Avizienis2012, Demis2015, Sillin2013, Stieg2012}. Simulated random memristor networks have been employed to implement {\em reservoir computing} (RC) \cite{Jaeger2001, Maass2002}. RC is a computational approach initially inspired by cortical microcircuits, in which computation takes place by translating the intrinsic dynamics of an excited medium, called a reservoir, into a desired output. By observing the nonlinear responses of a random resistive switch network to different input signals, it was possible to perform simple pattern classification \cite{Burger2013, Kulkarni2012} as well as computationally more demanding tasks by using multiple independent random assemblies \cite{Burger2015}. In contrast to the simulation-based results, work on {\em atomic switch networks} (ASN) has demonstrated physical random assemblies and discussed fabrication parameters that determine the network morphology \cite{Demis2015}. Based on these fundamental device and network characteristics that resemble the complexity of biological brains, it was argued that these networks are viable candidates for the physical realizations of brain-inspired information processing, such as reservoir computing.

% \note[JB]{4.Contribution of this work}\newline
Here we present a modeling and simulation framework that enables a detailed analysis of resistive switch network morphologies as determined by nanowire lengths, distributions and density. The computational capabilities of different network morphologies are analyzed with respect to the compressibility of the measured network signals. We put the computational capabilities into perspective by comparing to corresponding energy-consumption data. This comparison outlines a trade-off between computation and energy consumption. A modular approach is presented that provides a more energy-efficient architecture while achieving higher computational capabilities. Our results demonstrate what constitutes computationally useful networks with respect to network morphology, density and signal amplitudes. Based on the used modeling parameters, future fabrication of random resistive switch networks can be guided to achieve a desired trade-off between computational capacity and energy consumption.

\section{Materials and Methods}

% \note[JB]{In this section I'm thinking of addressing the material science community by discussing fundamentals of memristive devices and how it is grounded in the physics of the used materials. Here I'd build up from the device to the network modeling.} \newline

%%%%%%%%%%%%%%%%%%%%%%%%%%%%%%%%%%%%%%%%%%%%%%%%%%%%%%%%%%%%%%%%%%%%%%%%%%%%%%%%%%%%%%%%%%%%%%%%%%%%
%%%%%%%%%%%%%%%%%%%%%%%%%%%%%%%%%%%%%%%%%%%%%%%%%%%%%%%%%%%%%%%%%%%%%%%%%%%%%%%%%%%%%%%%%%%%%%%%%%%%
%%%%%%%%%%%%%%%%%%%%%%%%%%%%%%%%%%%%%%%%%%%%%%%%%%%%%%%%%%%%%%%%%%%%%%%%%%%%%%%%%%%%%%%%%%%%%%%%%%%%

\subsection{Memristive Devices and Atomic Switches}

%\note[JB]{As random fabrication has only been shown for ASN as as a gap-type device and not for the metal-insulator-metal memristors, we need to establish that the switching characteristics are similar and that our approach is valid for both types of devices. The functional similarity/equivalence of atomic switches and memristors was previously done by \cite{Hasegawa2012}. I hope we can find our main arguments in there. Current text is an older scratch and is not well structured and focused enough.} \newline

Memristive devices and atomic switches have been investigated in the context of random networks and reservoir computing. In this section we will establish the functional similarities and argue that the methods and results presented here are valid for both device types, and hence for a larger range of resistive switches. Detailed comparisons of memristive devices and atomic switches can be found in \cite{Chang2013, Hasegawa2011}.

The {\em metal-insulator-metal} (MIM) structure of the memristor, where the insulator is typically a metal-oxide such as TiO$_2$ or WO$_x$ \cite{Chang2013, Strukov2008}, establishes changes in the device conductance by redistributing oxygen vacancies within the metal-oxide. In the absence of an input bias, the oxygen vacancies remain at their positions, which leads to the history-dependent conductance of the device. Volatile behavior was demonstrated for a WO$_x$ memristive device \cite{Chang2011a}. Here spontaneous diffusion of the oxygen vacancies causes a dissolution of conducting channels in the WO$_x$ thin-film and a return to a low-conducting ground state of the device.

The atomic switch as a gap-type device based on crossing Ag$_2$S and Pt wires, achieves conductance changes by changing the concentration of Ag$^+$ cations which will allow growing metal protrusions of Ag atoms that eventually form a bridge between the two wires. The width of that bridge determines the conductance of the device \cite{Hasegawa2011}. In the absence of an input bias, the thermodynamically unstable atomic bridge eventually dissolves and the atomic switch returns to a low conductive equilibrium state \cite{Ohno2011}.

An important functional similarity between memristors and atomic switches is the nonlinear relation between the applied input and the resulting device state and current. Strukov and Williams presented an exponential ionic drift model that describes how an applied electric field changes the effective activation barrier and velocity for ionic-drift within the metal-oxide \cite{Strukov2009}. Similarly, Tamura {\em et al.} observed the exponential dependence of applied bias and switching time and argued this to be caused by a required minimum activation energy necessary to form a metal bridge within the atomic switch \cite{Tamura2006}.

Having established the qualitative similarities between memristive devices and atomic switches that are relevant to nonlinear computation, the rest of this paper will refer to both types of devices as resistive switches, allowing the presented concepts to be translated to specific device implementations.

\subsection{Resistive Switch Modeling}
\label{sec:deviceModel}

In this section we will describe the device model used in our simulations. Our focus is on capturing the fundamental switching characteristics of the discussed devices, not on precisely reproducing empirical data obtained from a specific device. As outlined in the preceding section, resistive switches are characterized by history-dependent nonlinear conductance changes and state decay.

We adopt a memristor model as presented for $WO_x$ devices in \cite{Chang2013, Chang2011}. In the original model the conductance as well as the state change are defined as follows:
\begin{align}
G = \left[\left(1-w\right)\epsilon\left[1-\exp\left(-\theta V\right)\right] + w \gamma \sinh\left(\delta V\right)\right] \frac{1}{V} 
\label{eq:current}
\end{align} \vspace{-4.5mm}
\begin{align}
\frac{dw}{dt} &= \lambda  \sinh\left(\eta V\right) - \frac{w}{\tau}
\label{eq:dw}
\end{align}

Here the internal device state is modeled by $w$, $V$ is the applied input bias, $\epsilon, \theta, \gamma, \delta, \lambda, \eta,\tau$ are the model parameters calculated from the experimental data.
%Note that we have increased $\lambda$ to lower the switching threshold. This was done as atomic switches show much lower thresholds, as well as to avoid the need for large voltages that would cause numerical instabilities inside the $\sinh$ term.
The nonlinear switching, as explained by the exponential ionic drift model \cite{Strukov2009}, is modelled by the $\sinh$ term. This model captures well the nonlinear switching as a function of the applied input and the current state.

However, such a first-order model, using only one variable to describe the device conductance, does not capture effects such as Ag$^+$ cation concentration changes. Such a change affects a device's response to future bias signals, but does not reflect into the actual device conductance. In \cite{Hasegawa2011} this difference was described as the memristor using a single variable to model the size of an ion-doped area, while the atomic switch uses two variables, one to model the height of an Ag protrusion, and another to model the width of the atomic bridge that emerges after the Ag protrusion has reached a sufficient height. Recently a second-order memristor model was presented that follows a similar modeling approach to the atomic switch \cite{Kim2015}. Here the second variable that functions as an enabler for a subsequent change in conductance is the internal device temperature. Application of an applied bias signal increases the internal device temperature due to Joule heating, which in turn affects drift and diffusion processes described earlier. 

To account for effects such as Joule heating or Ag protrusions, we use equation \ref{eq:dw} to model an internal state $w'$ that does not directly reflect in the device conductance. Furthermore, we extend the model to implement the different state decays based on the device state. As shown in \cite{Chang2011a, Ohno2011}, the rate of dissolution of the atomic bridge or the diffusion of ions to an equilibrium state is state-dependent and enables short- and long-term memory within a single device. We adopt equation \ref{eq:dw} as well as describe state variable $w$, which models the device conductance as:
\begin{align}
\frac{dw'}{dt} &= \lambda  \sinh\left(\eta V\right) - \frac{w'}{\tau}(1-w')
\label{eq:diw}
\end{align}
\begin{align}
w &= f(w')
\label{eq:f_hyst}
\end{align}

% We employ three different functions $f$ that map the internal state $iw$ to the external state $w$ in order to investigate the importance of state transitions to the computational approach. The first derivative of that model is a linear conversion from $iw$ to $w$ which effectively reduces the second-order model back to the first-order model. The second derivative applies rounding of $iw$ to $w$ and allows multi-state conductances without small changes in $iw$ directly reflecting into $w$. This implements a simplified model of the atomic switches and memristors with multiple states discussed in \cite{Chang2011a, Hasegawa2010}. 

For our simulations we employ a binary switching function $f$ that thresholds $w'$ to create two distinct conductances. Binary behavior of atomic switches was shown in \cite{Stieg2014} and is also found in some memristive devices \cite{Gaba2013}. Different levels of sub-surface Ag$^+$ concentrations are required to establish or dissolve an atomic bridge (length and width of $Ag$ protrusion) \cite{Ohno2012}. This implies different threshold values for the internal state variable $w'$ to perform device switching. We model this by applying a hysteresis function to $w'$.

As random assemblies of resistive switch networks will exhibit variations in nanowire diameters, and hence in the resulting device sizes (e.g., gap size of atomic switches), variation is an integral system characteristic \cite{Demis2015}. In \cite{Sillin2013} device parameter ranges for ASN were shown. These parameter ranges were not simply described by a Gaussian distribution around a mean value, but could span several orders of magnitude. For our experiments we apply a simplified model to create device variation where we draw device parameters uniformly from respective parameter ranges.

\subsection{Network Modeling}
\label{sec:network}

While the application of memristive networks to reservoir computing was all simulation based \cite{Burger2015, Burger2013, Kulkarni2012}, physical assemblies of atomic switch networks were reported in \cite{Avizienis2012, Demis2015, Sillin2013, Stieg2012, Stieg2014}. A simple modeling approach with focus on localized conductance changes of such networks was shown in the supplementary documentation of \cite{Sillin2013}.

Here we expand the modeling of such networks with the aim to investigate the relation between network morphologies, computational capabilities, and energy consumption. Network morphology is defined by the average length of nanowires. Their density can be controlled by the underlying copper seed posts. In \cite{Sillin2013} the effects of copper seed posts' shape, size, and pitch on the network's nanowire length and density were described. Smaller seed posts lead to long wires while larger seed posts lead to more fractal local structures. The pitch of the seed posts is a control parameter for the network density.

The connection of the random assembly of nanowires with the size of the seed posts implies that the distributions of the nanowire lengths can be described in terms of some probability density function. Large seed posts would cause more localized connections while small seed posts would result in long-range connections. Hence, we use a {\em probability density function} (PDF) that can capture different distributions. We use a beta-distribution for our purposes. A beta-distribution $B$ produces values in the interval of $[0,1]$ and is controlled by the parameters $\alpha$ and $\beta$ that define the mean value and the skewness around the mean.
\begin{equation}
    \begin{aligned}
        P(x,\alpha, \beta) = \frac{1}{B(\alpha,\beta)} x^{\alpha-1}(1-x)^{\beta-1}  \quad \text{(PDF)}\\
        \mu = \frac{\alpha}{\alpha + \beta} \quad\quad \text{(Mean value)} \\
        S = \frac{2(\beta-\alpha)\sqrt{\alpha+\beta+1}}{(\alpha+\beta+2)\sqrt{\alpha\beta}}  \quad \text{(Skewness)}        
    \end{aligned}
\end{equation}
$B(\alpha,\beta)=\int_0^1 x^{\alpha-1}(1-x)^{\beta-1}dx$ is a normalization factor ensuring that $P(x,\alpha, \beta)$ is a probability measure.
\begin{figure}
\def \w{0.45\textwidth}
\centering
\subfloat[Beta Distributions]{
\includegraphics[width=\w]{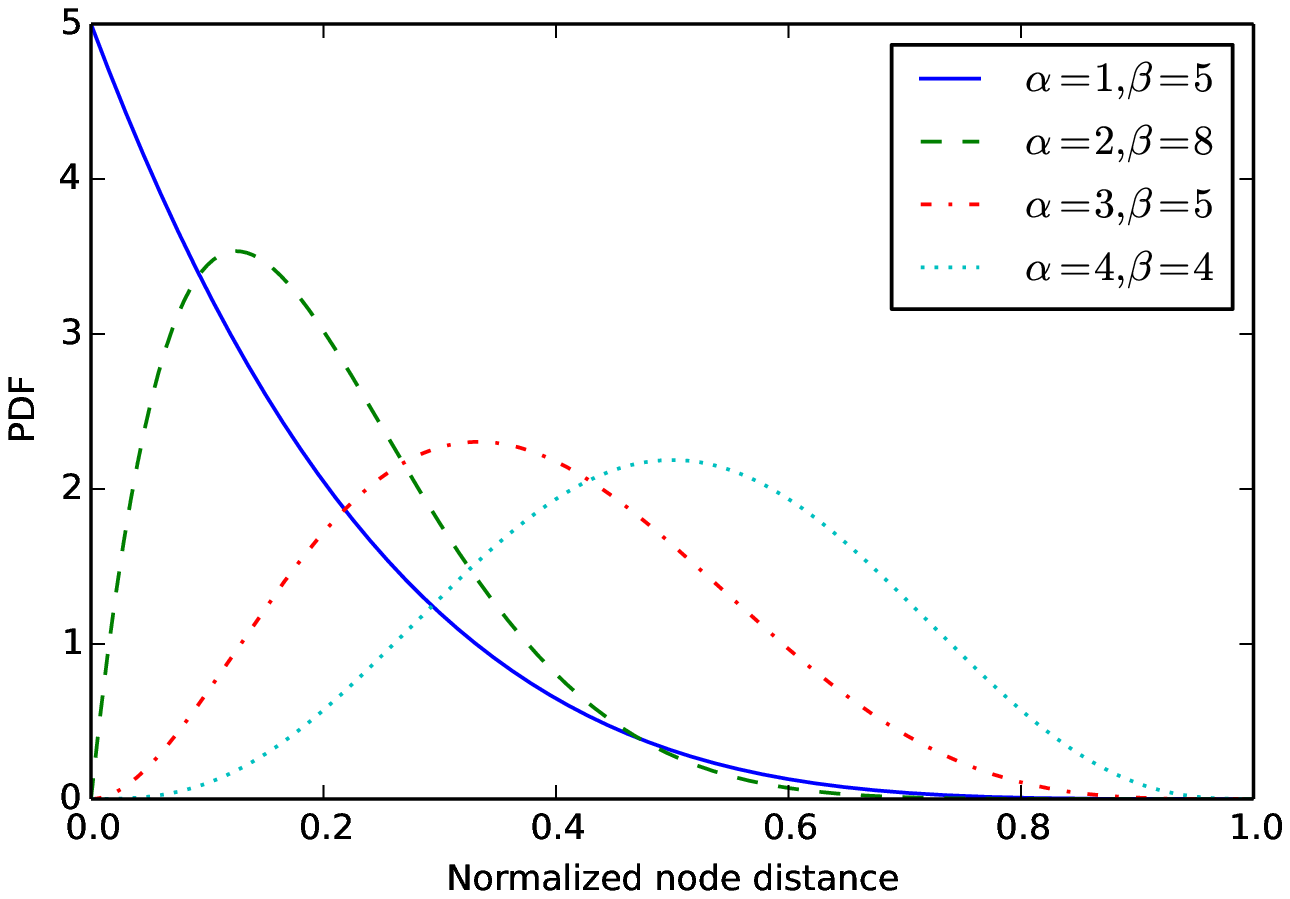}
\label{fig:beta}
}
\quad\quad\quad
\subfloat[Network Example]{
\includegraphics[width=0.3\textwidth]{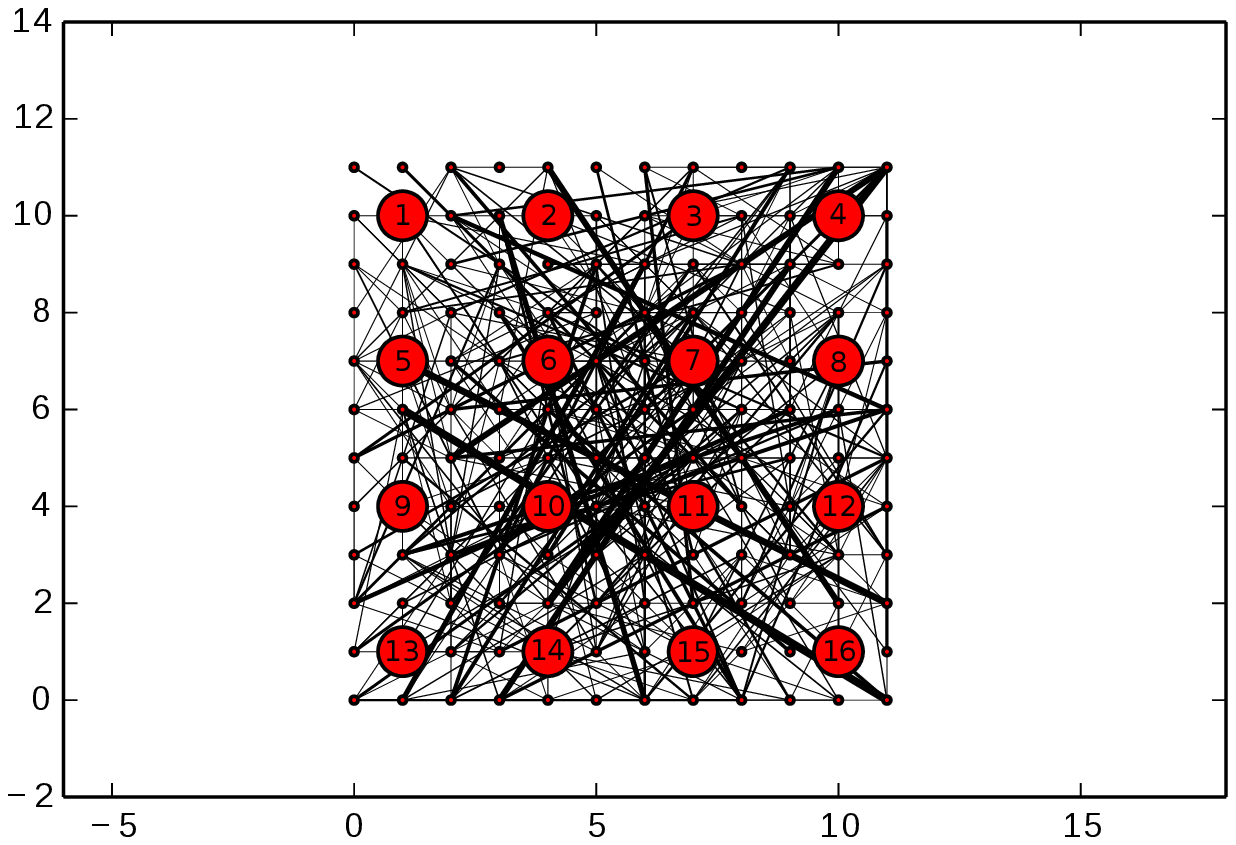}
\label{fig:network}
}
\caption{(a) Different beta distributions and the resulting relative nanowire lengths. The x-axis represents the normalized nanowire length and the y-axis the probability of creating a nanowire of that length. (b) Example graph of modelling a random $16_2$ memristive/atomic switch network with $\alpha=2, \beta=5,$ and $\xi=4$. The large circles represent the interface nodes to the underlying CMOS layer. The small circles create a narrower supporting grid that guides the density and morphology of the network structure. The graph edges represent memristive devices and the edge width is an indicator for the connectivity range with increasing edge widths representing longer-range connections between nodes.}
\label{fig:network_modeling}
\end{figure}
We use this PDF to model the distribution of the relative nanowire lengths within a random network. To translate the PDF into nanowire lengths we create a map with normalized Euclidean distances ($[0,1]$) from every seed post within the network. Starting from an initial node, we add a nanowire to the network by drawing a value from $P(x,\alpha, \beta)$ that determines the nanowire length. Beta-distributions with a focus on local connections ($\alpha < \beta$) will produce more fractal structures as nanowires branch out around the selected starting node for the wire. Distributions favoring long-range connections ($\alpha > \beta$) will connect distant parts of the underlying grid of seed posts. In Fig. \ref{fig:beta} we show some examples of $P(x,\alpha, \beta)$ as a function of the control parameters $\alpha$ and $\beta$.

% \begin{figure}
% \begin{center}
% \includegraphics[width=0.6\textwidth]{grid_network.eps}
% \caption{Example graph of modelling random memristive/atomic switch networks. The large circles represent the interface nodes to the underlying CMOS layer. The small circles create a narrower supporting grid that guides the density and morphology of the network structure. The graph edges represent memristive devices and the edge width is an indicator for the connectivity range with increasing edge widths representing longer range connections between nodes.}
% \label{fig:network}
% \end{center}
% \end{figure}

For the modeling of the underlying seed post grid we define two post types. Interface seed posts allow the application and reading of network voltages. These posts are established at a more coarse-grained pitch. The second grid is a supporting grid and more fine-grained than the interface grid. This supporting grid improves the formation of fractal structures for localized wire growth \cite{Demis2015} and the establishment of more complex structures  between the interface nodes; without a supporting grid  establishment of multiple devices between two interface nodes will not be favourable which limits the morphological diversity. The density of a network is defined by the indegree $\xi$ of a node. $\xi$ defines the average number of devices connected to each grid node. The number of resulting devices, and hence the density, is then given by the total number nodes in the supporting grid multiplied by $\xi$. The fabrication of networks with different densities was demonstrated in \cite{Sillin2013}. In Fig. \ref{fig:network} we show an example network model with 16 interface nodes, a supporting grid at a third of the pitch of the interface grid, and a mix of short and long-range connections. For all simulations we have used 16 interface nodes arranged in a $4\times4$ grid as shown in Fig. \ref{fig:network}. If not mentioned otherwise we have used a supporting grid at half the pitch of the interface grid. Throughout this paper we will refer to this network as of size $16_1$, meaning that it has $16$ interface nodes and $1$ node  between two interface nodes (half the pitch).

\begin{table}
\begin{center}
\caption{Examples of network parameters}
\begin{tabular}{cccccc} \hline

ID & $\alpha$ & $\beta$ & $\xi$ & Connection length & Density \\\hline
N1 & 1 & 1 & 2 & uniform & sparse \\
N2 & 1 & 10 & 8 & short & dense \\
N3 & 10 & 1 & 4 & long & semi-sparse \\
N4 & 10 & 10 & 6 & medium (small $\sigma$) & semi-sparse \\
N5 & 5 & 5 & 2 & medium (medium $\sigma$) & sparse \\\hline

\end{tabular}
\label{tab:networks}
\end{center}
\end{table}

For a more intuitive understanding of how the defined network parameters inter-relate, we give five examples and describe the resulting network morphology (Table \ref{tab:networks}). In Fig.~\ref{fig:entropy} and \ref{fig:entropy_v}, we added the network IDs to the presented results  to  illustrate the relation between network morphology and information processing capacities.

We simulate the resistive switch networks by treating them as temporarily stationary resistive networks that can be solved efficiently using the {\em modified nodal analysis} (MNA) algorithm \cite{Litovski1997}. After calculating one time step using the MNA, we update the memristive devices based on the node voltages present in the network to account for the dynamic state changes of memristors.

\section{Results}

\subsection{Network Morphology and Computational Capacity}
\label{sec:entropy}

As was outlined by Demis  {\em et al.} \cite{Demis2015}, the structural similarities of ASN and biological brains (i.e., fractal branching similar to dendritic trees) suggests that such complex random assemblies provide hardware platforms for efficient brain-inspired computing. A characteristic feature of cognitive architectures is the nonlinear transformation of an input signal into a high-dimensional representation more suitable for information processing \cite{Rabinovich2008,Buonomano2009}. In particular, for $N>M$, the input signal $\mathbf{u} \in \mathcal{R}^M$ is transformed to $\mathbf{x} \in \mathcal{R}^N$ by the dynamics of the cortical microcircuits driven by sensory inputs. In the case of random resistive switch networks this means that the measured signals at the interface nodes are ideally nonlinearly related to the input signal and then provide a platform for brain-inspired computing.

% AG: what does richness tell us
A suitable transformation of the input requires rich dynamics that can preserve relevant distinctions between different input signals in the high-dimensional space \cite{Maass2002}. This property is attributed  to the dynamics of a system in the critical regime where the distinctions between states do not diverge or converge \cite{Bertschinger:2004p1450,Langton:1990p954,PhysRevE.87.042808}. To provide such a rich dynamics, the activity of the nodes should show the least amount of redundancy. In other words,  the dynamics of the nodes should be as uncorrelated as possible. The question is how could one measure that and how this measure would be related to the parameters of the system.  

Here we introduce a simple measure that can describe the dynamics of random network signals in a way that is meaningful in the context of information processing. We study the compressibility of the network dynamics as a proxy for its richness. Specifically, we use {\em principal component analysis} (PCA) to transform the system dynamics into a principal component space \cite{bishop06prml}.

The distribution of variations in the principal component space indicates the amount of redundancy between the different dimensions of the original system. The variation in each principal component is given by the corresponding normalized Eigenvalue of the covariance matrix of the system. To calculate these, we record the network dynamics on all interface nodes as a result of an applied network input. All network signals are expressed in the network state matrix $X$. The covariance matrix of the network dynamics is then given by $\mathcal{C} = X^TX$, and the Eigenvalues are obtained by diagonalizing,  $\mathcal{C}=U \Lambda U^{-1}$. The diagonal elements of $\Lambda$, $\{\Lambda_1,\Lambda_2,\dots,\Lambda_N\}$, are the Eigenvalues of the corresponding dimension $i$ and can be normalized as $\lambda_i = \frac{\Lambda_i}{\sum_{i=1}^N \Lambda_i}$. Since $\lambda_i$ are normalized as a probability measure, we can describe their evenness using a single number $\mathcal{H}= - \sum_{i=1}^N \lambda_i \log_2(\lambda_i)$. This is an entropy measure and describes how evenly the $\lambda_i$ are distributed. In one extreme case, where the system nodes are all maximally correlated, only one Eigenvalue will be 1 and the rest will be zero, and the resulting entropy will be $\mathcal{H}=0$. In the other extreme case, where the nodes are maximally uncorrelated, the Eigenvalues will be identical and equal to $\frac{1}{N}$, and the resulting entropy will be $\mathcal{H}= \log_2 N$. In the latter case, every node in the network represents something unique about the properties of the input signal that cannot be described by a combination of the rest of the nodes.

\begin{figure}
\begin{center}
\includegraphics[width=0.75\textwidth]{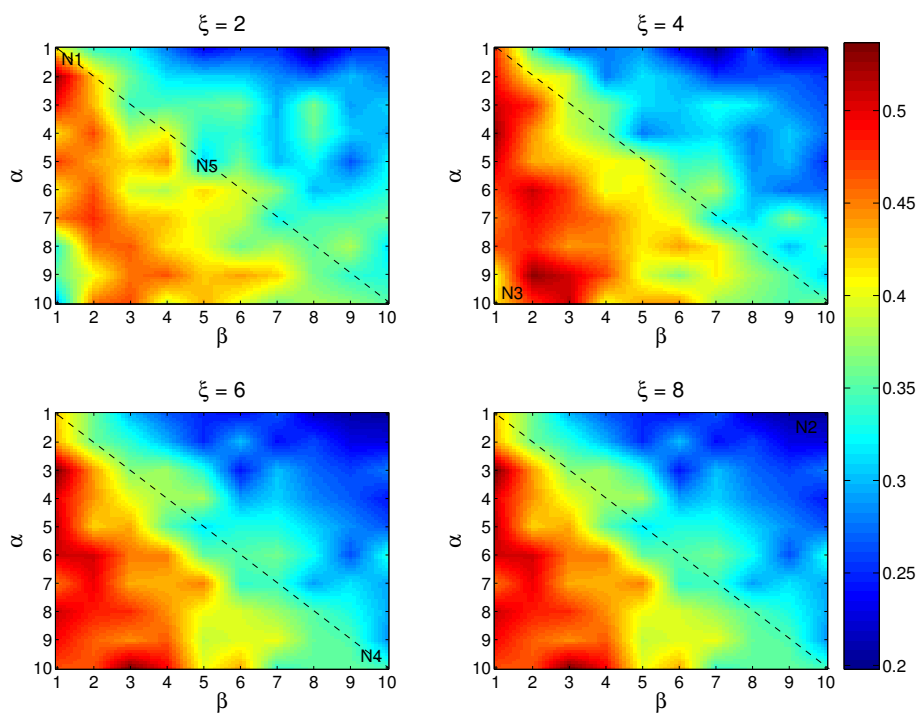}
\caption{Entropy as a function of the network topology defined by the beta distribution values $\alpha, \beta$ and the indegree $\xi$ ($v=8V$). Independent of the density, best entropies are achieved for more globally connected network morphologies described by $\alpha \le \beta$ (lower triangles of the plots). With respect to the network density an indegree of $\xi=6$ has resulted in best entropies. Lower $\xi$ values provide fewer signal paths and hence less network activity. Average numbers of network devices were $150, 250, 350, 450$ for $\xi=2,4,6,8$, respectively. Markers Nx refer to table \ref{tab:networks}.}
\label{fig:entropy}
\end{center}
\end{figure}

In the following we present entropy measurements as a function of the network morphology. As outlined in section \ref{sec:network}, the morphology is modeled based on a beta distribution with parameters $\alpha$ and $\beta$ as well as the indegree $\xi$ that defines the device density. We apply a 5Hz sine wave to the upper left interface node of a $16_1$ network and connect the lower right node to ground (0V). Fig.~\ref{fig:entropy} shows the network entropies as a function of $\alpha, \beta$, and $\xi$. Across all densities, the highest entropies, and hence the least linearly-dependent network states are achieved with a majority of the nanowires being equal or longer than half the normalized maximum euclidean distance (lower left triangles in Fig.~\ref{fig:entropy} where $\alpha \leq \beta$). Long-range connections spatially distribute the input signal across the network without much voltage drop. Hence, different areas of the network experience a sufficient bias voltage to exhibit switching dynamics useful to information processing. Increasing network densities, which in other words describes the number of devices per area, creates more signal paths and due to device parallelism also higher conductive connections. This also leads to better distribution of the input signal and to an expansion of the morphologies for which larger entropies can be achieved.

\begin{figure}
\begin{center}
\includegraphics[width=0.75\textwidth]{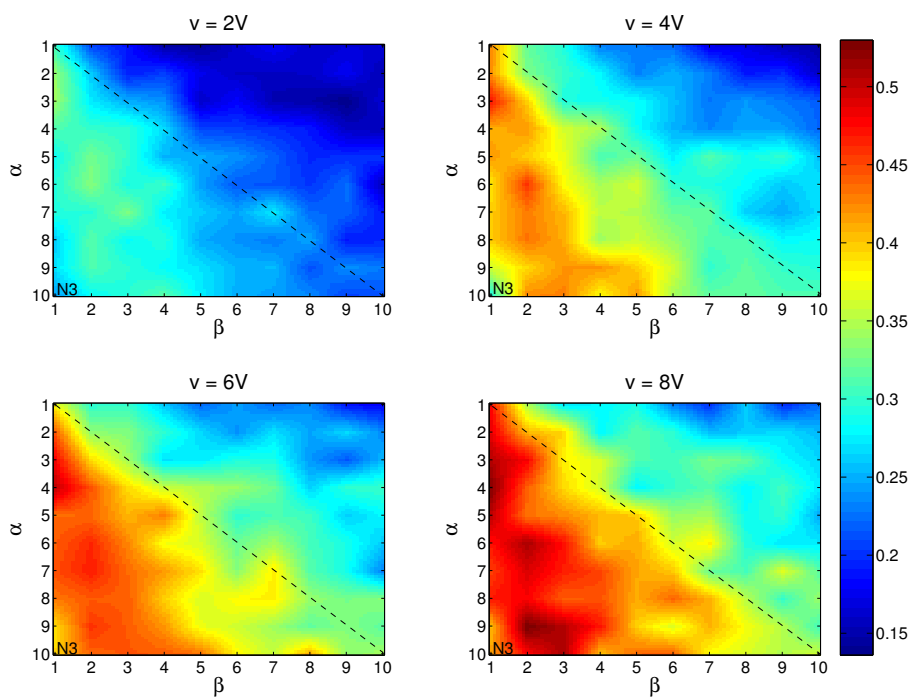}
\caption{Entropy as a function of the network topology defined by the beta distribution values $\alpha, \beta$ and the input signal amplitude $v$ ($\xi=4$). Increasing signal amplitudes $v$ lead to higher bias voltages for individual network devices and results in higher switching activity. Highest entropies are again achieved for more globally connected networks with $\alpha \le \beta$ (lower triangles of the plots). Markers Nx refer to table \ref{tab:networks}.}
\label{fig:entropy_v}
\end{center}
\end{figure}

Besides the network morphology, the amplitude $v$ of the input signal also greatly affects the network dynamics due to the exponential dependence of applied bias and either activation energy to form an atomic bridge (for atomic switches) or the velocity of ionic drift (for memristive devices). In Fig.~\ref{fig:entropy_v} we show the entropy as a function of $\alpha, \beta$, and $v$ for networks with an indegree $\xi=6$. It can be seen that the average entropy increases as we increase $v$. This is related to larger voltages enabling more devices to exhibit switching activity and hence affect the network dynamics. The similarity in the plots as compared to Fig.~\ref{fig:entropy} implies that increased input voltages also allow better spatial distribution of the input and more areas of the network to exhibit switching activity. Note that these plots present qualitative results on the dependence of $v$ and entropy, but the absolute values of $v$ are device dependent and can change for devices with different threshold behavior.

%The most recent publication on atomic switch networks \cite{Demis2015} has given further insights into the morphology of ASN and how it structurally resembles the complexity of biological brains. Hence, natural computing is the most promising application area for such architectures. Natural computing relies on the nonlinear mapping of an input signal into a higher dimensionality from which relevant features can be extracted. What this means, was shown qualitatively by \cite{Demis2015}. An applied input signal (i.e. Gaussian pulse) experienced nonlinear transformations due to the switching activity in the network. The collection of voltages measured across different interface nodes represented the signal at a higher dimensionality.

We also studied network sizes $16_0$, and $16_2$. $16_0$ networks have shown very similar results to the $16_1$ networks presented here. The morphological similarity between $16_0$ and more globally connected $16_1$ networks is that they do not form many local fractal structures. Contrary, the $16_2$ networks are characterized by more complex morphologies between the interface nodes. While this might closely resemble complex structures in biological brains, in the context of passive electrical networks, these fine-grained fractal structures with many devices in-between interface nodes lead to alleviated individual switching dynamics.
% \change[AG]{alleviated individual switching dynamics}{reduced switching activity}. 
While this could be circumvented in simulations by increasing the input amplitude $v$, in practice this is not a viable approach for reasons of safe operation and energy consumption.

\subsection{Energy Consumption}
\label{sec:energy}

\begin{figure}
\def \w{0.4\textwidth}
\centering
\subfloat[Average Energy Consumption]{
\includegraphics[width=\w]{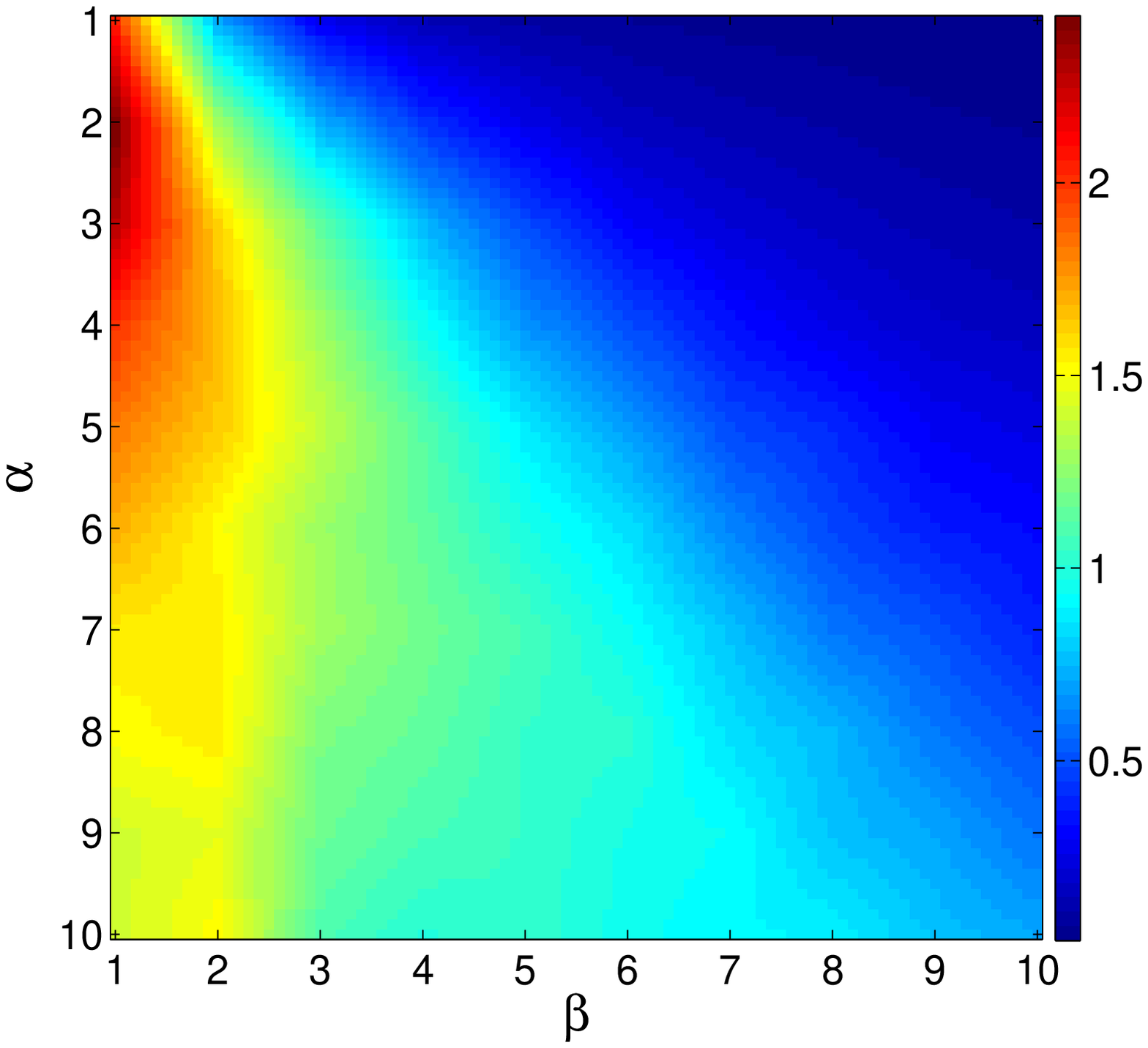}
\label{fig:ener_sum}
}
\subfloat[Energy vs. Entropy]{
\includegraphics[width=\w]{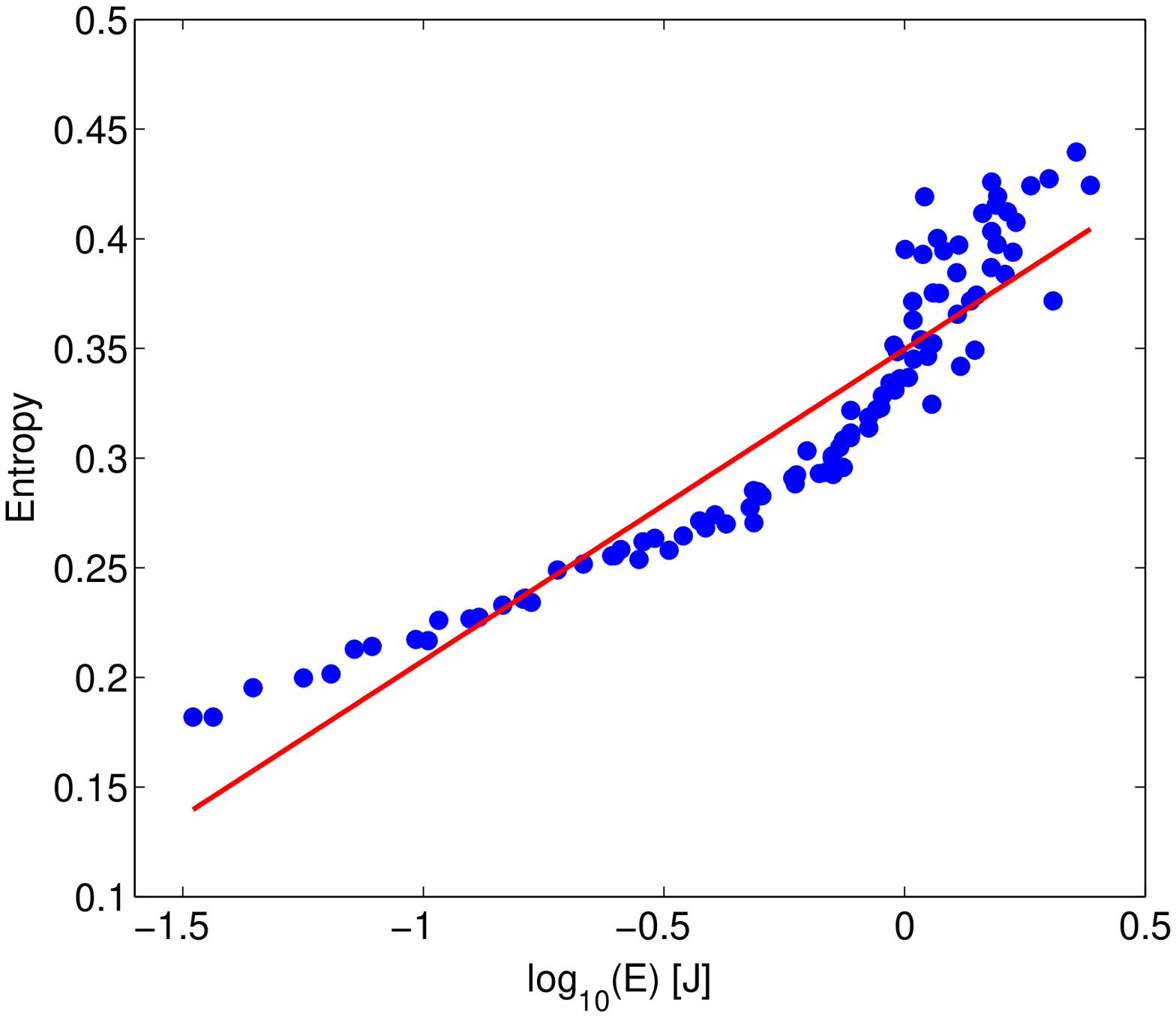}
\label{fig:ener_entr_scatter}
}
\caption{Energy consumption in random resistive switch networks. (a) Averaged energy consumption for different network morphologies expressed as $log_{10}(E)$. Energy grows exponentially when transitioning from locally connected ($\alpha=1, \beta=10$) to more globally connected networks ($\alpha \le \beta$). The link between energy and entropy is shown in (b). Entropy grows linearly with exponential energy increase.}
\label{fig:ener}
\end{figure}

In the previous section we have shown that best computational capabilities are achieved for dense, globally connected networks ($\alpha \le \beta$) and larger signal amplitudes $v$. The viability of these results has to be evaluated with respect to the energy that would be consumed by such networks. Based on the application of a $5$\,Hz sine wave with amplitude $v$, we calculate the total energy consumption of a network over time $T$ as $E(t) = \int_{i=0}^T V_i(t)I_i(t)dt$, with $V(t)$ being the time-dependent signal amplitude and $I(t)$ the current drawn by the network. As the energy consumption is very application-specific, we consider our findings as qualitative measures that highlight the general relations between the network parameters and the consumed energy. Absolute energy numbers presented here are not of relevance, only the information on how drastically energy consumption changes with network parameters. Fig.~\ref{fig:ener_sum} shows the resulting energy consumption for different $\alpha$ and $\beta$, averaged over different $\xi$ and $v$. The distributions of the energy data across the $\alpha$ and $\beta$ plane resemble the entropy distributions seen in Fig.~\ref{fig:entropy} and \ref{fig:entropy_v}, which confirms that high entropies come at the cost of high energy consumption (relative to the consumption at low entropies) caused by high conducting paths in denser networks. This finding is further supported by plotting energy vs. entropy (Fig. \ref{fig:ener_entr_scatter}). Here we plot the averaged energy against the averaged entropy for corresponding setups. We can see how entropy grows with energy. However, as the energy grows exponentially, increasing entropy can lead to over-proportional energy consumption.

\subsection{Improved Information Processing with Random Resistive Switch Networks}

As the computational performance of a single random network is limited by exponentially increasing energy consumptions or strong linear dependence at low energies, we will outline an approach to increase entropy with linear growth of energy. In \cite{Stieg2014} a hierarchical approach of ASN was presented that combined multiple small-world networks on a single chip. Similarly, we have shown a hierarchical approach that embedded independent networks (similar sizes as presented here) in a reservoir computing architecture and showed that a memory and computationally demanding application could be solved \cite{Burger2015}.

\begin{figure}
\def \w{0.3\textwidth}
\centering
\subfloat[Network Example]{
\includegraphics[width=0.2\textwidth]{grid_network_square.eps}
\label{fig:grid_network_comp}
}
\subfloat[Single Network]{
\includegraphics[width=\w]{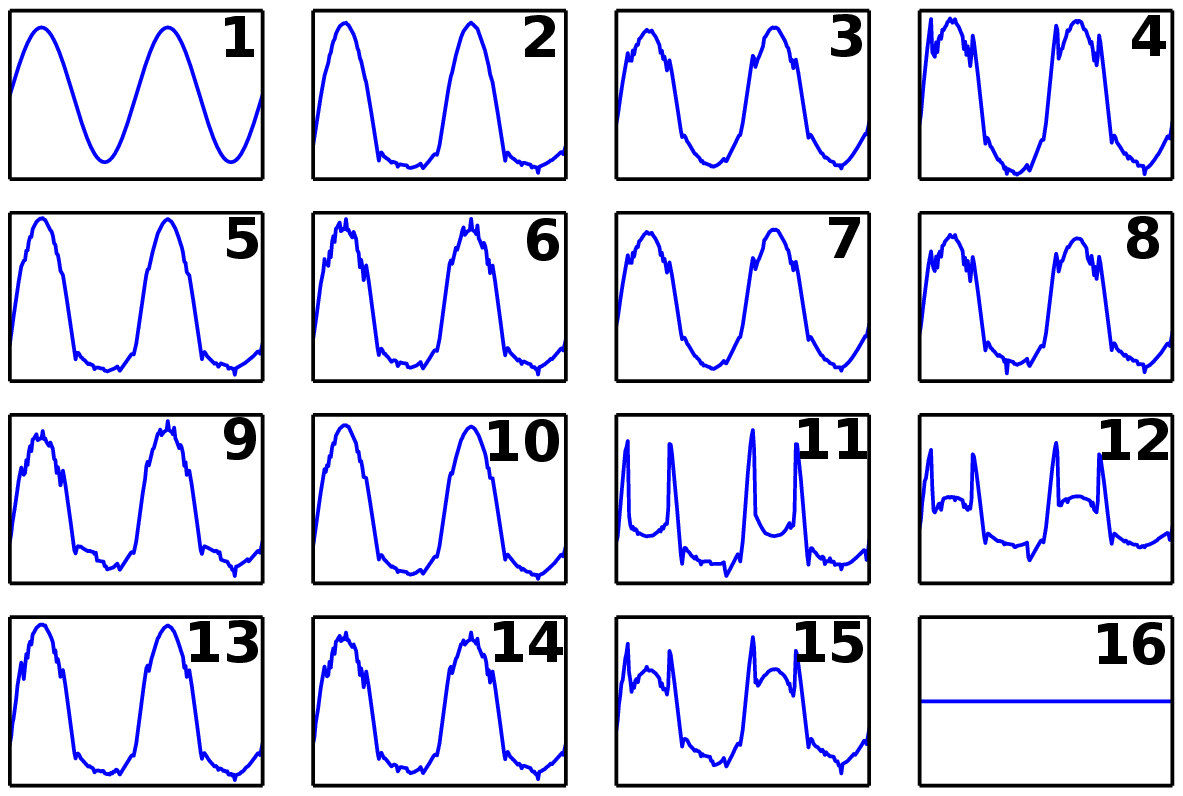}
\label{fig:ASN_voltages}
}
\subfloat[Independent Networks]{
\includegraphics[height=97pt,width=\w]{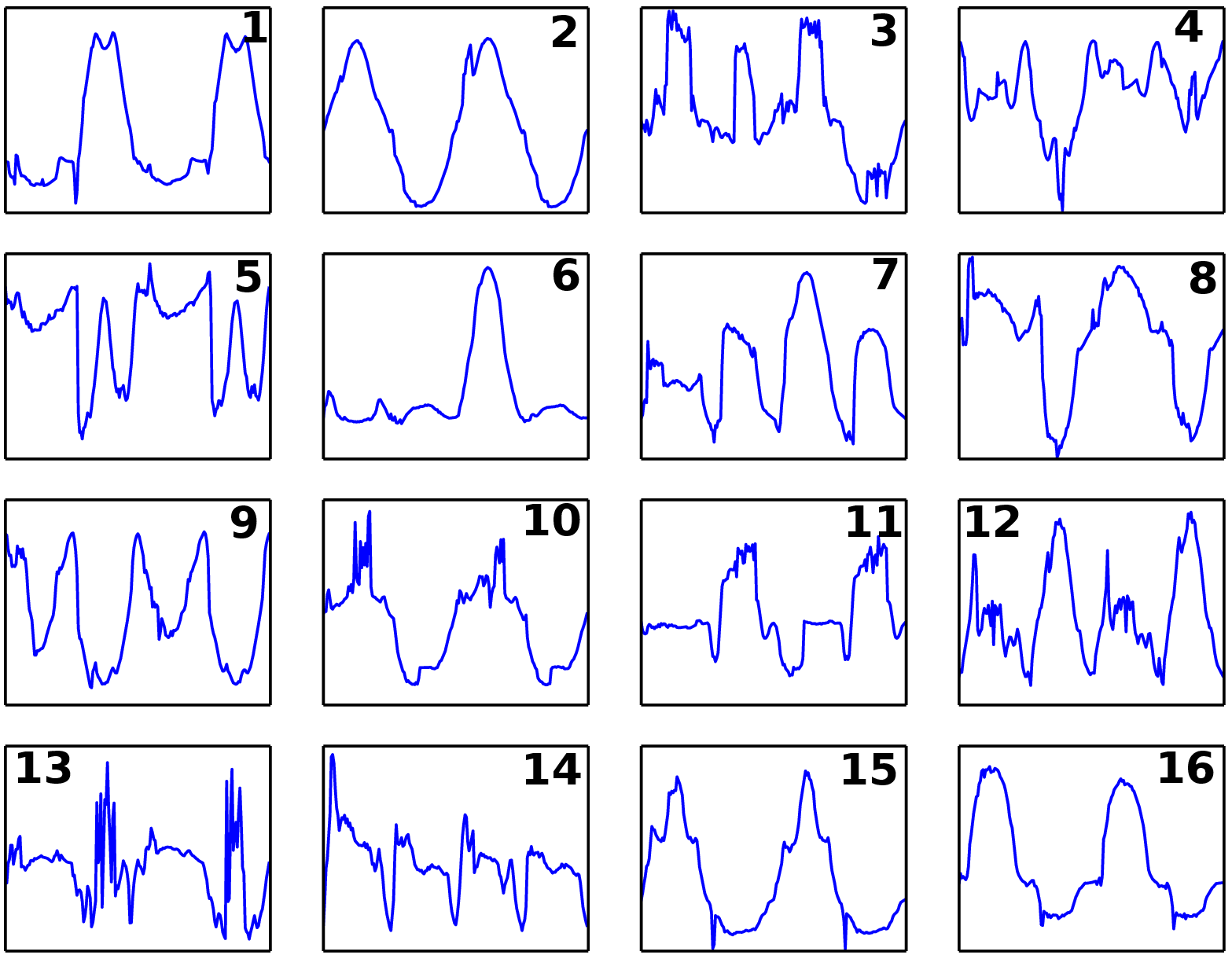}
\label{fig:SCR_voltages}
}
\caption{Examples of networks and node signals. (a) Example $16_2$ network with $\alpha=2, \beta=5,$ and $\xi=4$ indicating the 16 interface nodes. (b) 16 Signals measured from a single $16_1$ network with the input applied to node 1 and the ground node connected to node 16. The presented signals have an entropy of $0.37$. The numbers correspond to the physical location of the nodes within the network. (c) Signals measured from 16 independent $16_1$ networks. The entropy for this example is $1.79$. Signals were measured as the difference between interface nodes 2 and 9. The numbers represent the network number. All networks were created with $\alpha=1, \beta=5, \xi=4, v=2V$}
\label{fig:16_voltages}
\end{figure}

\begin{figure}
\begin{center}
\includegraphics[width=0.75\textwidth]{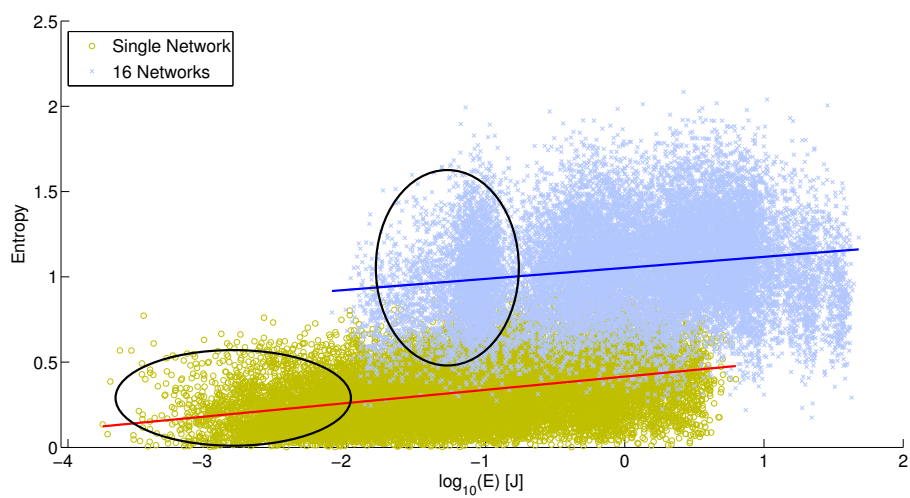}
\caption{Energy vs. entropy for single networks and a hierarchical approach with 16 independent networks. The circled areas highlight an example of networks with the same network parameters ($v=2, \xi=2$). Scaling up the number of independent networks is more energy- as well as computationally-efficient as compared to increasing the entropy of a single network by means of scaling up either density or voltage. Data obtained with $v=2$ and $\xi=2\dots8$.}
\label{fig:ener_entr_scatter_2class}
\end{center}
\end{figure}

The  concept relies on extracting only a subset of signals from each independent network. This allows  harnessing the different processing caused by the different random structures of the independent networks. Furthermore, in \cite{Burger2015} we have extracted a differential signal from each network. By retrieving a network output as the difference of two network nodes, differences in the spatial and temporal dynamics within networks can be better captured than using signals measured with respect to a common ground. In Fig.~\ref{fig:16_voltages} we show two examples of 16 network signals. Fig.~\ref{fig:ASN_voltages} shows signals obtained from a single random network (qualitatively comparable to signals presented in \cite{Demis2015}). Readouts from  a single network show some nonlinearities, however, it is mostly characterized by strong linear dependence across the signals (low entropy of $0.37$). In contrast, 16 independent networks exposed to the same input, we can significantly increase the richness of the measured signals. In this example the entropy is $1.79$. The total energy consumption of the hierarchical independent networks scales linearly with the number of networks. Considering the gain in entropy, this approach poses a much more viable approach for brain-inspired information processing than relying on single networks with high density and high signal amplitudes. A comparison of how energy and entropy relate in the two presented setups is shown in Fig. \ref{fig:ener_entr_scatter_2class}. The plotted data represents collected data from single networks as well as for the 16 networks approach with $\xi=4$ and increasing $v$. The circled areas approximately mark the energy and entropy data obtained with a $2$\,V input signal. The average energy difference is $16$ which corresponds to the number of networks used. However, compared with single networks that exhibit similar energy consumption (around $v=4..6$) significantly higher entropies can be achieved.

%%%%%%%%%%%%%%%%%%%%%%%%%%%%%%%%%%%%%%%%%%%%%%%%%%%%%%%%%%%%%%%%%%%

% \begin{figure}
% \begin{center}
% \includegraphics[width=0.75\textwidth]{entr_per_ener_2volt_ASN_SCR.eps}
% \caption{Entropy per energy compared between single network and 16 networks. }
% \label{fig:ener_per_entr_asn_scr}
% \end{center}
% \end{figure} 

% \begin{figure}
% \begin{center}
% \includegraphics[width=0.75\textwidth]{ener_entr_scatter_SCR_all.eps}
% \caption{Scatter plot SCR}
% \label{fig:ener_entr_scatter}
% \end{center}
% \end{figure} 

% \begin{figure}
% \begin{center}
% \includegraphics[width=0.75\textwidth]{ener_entr_scatter_ASN_SCR.eps}
% \caption{Scatter ASN SCR}
% \label{fig:ener_entr_scatter_asn_scr}
% \end{center}
% \end{figure} 

\section{Discussion}

% We have investigated random resistive switch networks and outlined the connections between network morphology, density, and input amplitude. By means of measuring the compressibility of the measured network signals we have evaluated the entropy as an indicator of the computational capabilities. While higher entropies come at the expense of exponentially increasing energy consumptions, the hierarchical approach of utilizing multiple small-world networks has presented a trade-off with larger entropy at reasonable energy consumption.

% The presented entropy measures were average values based on 50 independent simulations for each data point. For larger entropies above $0.7$ we have observed average standard deviations of $\sigma=0.3$. These high $\sigma$ can be related to the random network morphology. At this point we cannot say if such large variations would be observed in physical networks or if it is an artifact of our modeling approach. Future research will further investigate the network morphology in an attempt to develop a more detailed understanding of computationally useful network morphologies.

We have presented a detailed analysis of how the morphology of random resistive switch networks affects computational capacity and energy consumption. It will be an application-specific choice of how much entropy is required by the individual networks and what costs one is willing to pay in terms of energy. The hierarchical approach and the better entropy per energy ratio is possible as we utilized different independent networks. This implies that cognitive computing is not merely the product of sufficient excitement of network elements but is also rooted in the heterogeneous morphologies across networks, as demonstrated for neural populations in cortical microcircuits \cite{Honey2010}. This ability to harness randomness is a strong argument for the future fabrication of nanoscale resistive switch networks. We found that networks with strong fractal structures (i.e., $16_2$ networks) performed worse. However, in biological brains such fractal structures (dendrites) are an integral part of the information processing. Biological dendrites are active components with voltage- and calcium-gated ion channels that regulate synaptic responses \cite{Shah2010}, while the fractal structures in resistive switch networks are purely passive elements. Further investigation of the functional differences might provide better fabrication techniques and methods to utilize fractal resistive switch networks.

% Optimize information gain of single network - differential reading

% \note[AG]{Maybe in discussion (if we have to) we should give numbers about high level message and not low level details}.

\section{Conclusion}

% \remove[AG]{We have presented an in depth discussion of design parameters and computational capacities of nanoscale resistive switch networks in the context of brain-inspired information processing.} 
The nonlinear dependence of input bias and state transitions in resistive switches allows random assemblies of such devices to exhibit complex behavior useful to computation. The design parameters used to control the simulated network morphologies relate to design parameters for the physical fabrication of the networks. We have shown how network morphologies correlate with computational capacities and energy consumption. A hierarchical approach was presented that allows us to increase computational capacities more energy-efficiently as compared to increasing information processing in single networks. These findings provide insights in the abilities and limitations of random nanoscale resistive switch networks and should serve as a design guide for  investigation and fabrication of future nanoscale computing architectures.

\section*{Acknowledgments}
This work was supported by the National Science Foundation under awards  \# 1028378, \# 1028238, and \# 1318833, and by DARPA under award \# HR0011-13-2-0015. The views expressed are those of the author(s) and do not reflect the official policy or position of the Department of Defense or the U.S. Government. Approved for Public Release, Distribution Unlimited.

% \section*{Conflict of Interest}

\bibliographystyle{AIMS}
\bibliography{bibliography}

\end{document}